\documentclass{article}

\usepackage{microtype}
\usepackage{graphicx}
\usepackage{booktabs}
\usepackage{array}
\usepackage{tabularx}
\usepackage{xurl}
\usepackage{hyperref}

\usepackage[preprint]{icml2026}
\makeatletter
\renewcommand{\Notice@String}{}
\makeatother
\usepackage{amsmath,amssymb,mathtools}
\usepackage{cuted}

\newcommand{\Clocal}{C_{\mathrm{local}}}
\newcommand{\Cthroughput}{C_{\mathrm{throughput}}}
\newcommand{\Cquality}{C_{\mathrm{quality}}}
\newcommand{\etaH}{\eta_{H}}
\newcommand{\etacomp}{\eta_{\mathrm{comp}}}
\newcommand{\etarep}{\eta_{\mathrm{rep}}}
\newcommand{\etaact}{\eta_{\mathrm{act}}}
\newcommand{\etalocal}{\eta}
\newcommand{\chip}[2]{#1$\times$ #2}
\newcommand{\flop}[1]{$10^{#1}$ FLOP}
\newcommand{\scinum}[2]{$#1\times 10^{#2}$}

\icmltitlerunning{Does Distributed Training Undermine Compute Governance?}

\begin{document}

\twocolumn[
  \icmltitle{Does Distributed Training Undermine Compute Governance?}

  \begin{icmlauthorlist}
    \icmlauthor{Robi Rahman}{miri}
  \end{icmlauthorlist}

  \icmlaffiliation{miri}{Machine Intelligence Research Institute}
  \icmlcorrespondingauthor{Robi Rahman}{robi@intelligence.org}
  \icmlkeywords{AI governance, compute governance, distributed training}

  \vskip 0.3in
]

\printAffiliationsAndNotice{}

\begin{abstract}
Compute governance proposals often rely on the assumption that frontier AI training requires large, detectable computing clusters. However, recent advances in distributed training algorithms could allow developers to conduct frontier-scale training on distributed agglomerations of hardware, rather than needing large datacenter facilities. Developers who prefer not to be constrained by regulations may structure their hardware in a manner that evades the registration and monitoring requirements associated with compute governance. Therefore, regulations must be designed to detect and prevent illicit distributed training operations. This paper evaluates the feasibility of such evasion and outlines recommended countermeasures, including whistleblowing, chip tracking, forensic accounting, and memory and compute thresholds for clusters.
\end{abstract}

\section{Introduction}
\label{sec:introduction}

Many existing and proposed compute governance measures rely on the governing body's awareness of large agglomerations of computing hardware, known as compute clusters or supercomputers \citep{pilz2025supercomputers}. Training frontier models requires very large amounts of computing power, and frontier model developers prefer to concentrate the requisite hardware into large clusters. This configuration enables all the constituent chips to transfer data to the others over high-bandwidth interconnect. It also allows personnel to perform maintenance on all the hardware in a single location, and enables chips to share memory, which improves resilience to hardware malfunctions and reduces the requisite frequency of checkpointing \citep{erben2024failures}.

Many regulations based on compute have been proposed, and several have been enacted. US Executive Order 14110, in effect from October 2023 to January 2025, previously required developers conducting training runs larger than \flop{26} (or larger than \flop{23} if training primarily on biological sequence data) to report them to the government and disclose their cybersecurity and risk management practices \citep{whitehouse2023eo14110}. Articles 51 and 55 of the EU AI Act require general-purpose AI models trained with more than \flop{25} to undergo safety evaluations and red-teaming \citep{euaiact2024}. California SB 53 regulates developers who train models with more than \flop{26}, requiring transparency and incident reporting, prohibiting false or misleading statements about catastrophic risk posed by the models, and establishing penalties of up to \$1 million for violations \citep{california2025sb53}. \citet{scher2025asi} propose an international ban on model training above \flop{24} and fine-tuning above \flop{23}. \citet{baker2025verification} recommend measures to verify compliance with requirements such as compute usage limits, auditing of large training runs, and prohibitions on certain categories of workloads.

Enforcement of these regulations would be rendered ineffective if the regulated developers could hide their computing hardware. Frontier-scale datacenters cannot be hidden because their power consumption and physical footprints are too large, so they can be tracked with power grid monitoring and satellite imagery \citep{epoch2026datacenters}. However, recent advancements in distributed computing methods for ML model training have increased the feasibility of harnessing diffuse collections of hardware to perform large-scale training runs. The \emph{DiLoCo} (distributed low-communication training) family of algorithms compresses the gradients that are transferred between processors during training, theoretically allowing large-scale model training to be done with less than 100 Mbps of bandwidth, thousands of times less than the 400+ Gbps connections that are standard in modern datacenters \citep{ieee2017ethernet}. If a frontier-scale cluster were split into nodes\footnote{A node refers to a single, self-contained unit of computing hardware that is part of a larger, geographically dispersed network. In the scenarios modeled, we investigate the feasibility of training with nodes smaller than the hardware reporting threshold proposed in \citet{scher2025asi}.} smaller than the regulator's monitoring threshold, it would be undetectable by thermal and electrical means, obligating enforcement to rely on network monitoring and in-person observations. Furthermore, once a node is identified, it is not certain that the regulator would be able to prove that it is part of a distributed training operation.

This work develops a comprehensive efficiency model for distributed training based on the state of the art as of 2026, evaluates the feasibility of large-scale training runs using a variety of types and sizes of processors, and assesses the efficacy of countermeasures for detecting regulatory evasion. We recommend that regulations governing models be accompanied by registration requirements for clusters exceeding compute throughput and/or accelerator memory thresholds, in order to prevent evasion. An interactive simulator is published online\footnote{Simulator available at \url{https://intelligence.org/research/distributed-training-simulator}; source code available at \url{https://github.com/robirahman/miri-decentralized-training-report}.} to assist policymakers and governance researchers in both understanding the problem and designing standards that are robust to such evasion.

\section{Background and Related Work}
\label{sec:background}

\citet{stich2019local} introduced \emph{local SGD}, a variant of stochastic gradient descent that maintains copies of a machine learning model on multiple different processors, training each copy independently for $H$ steps before averaging parameters, which reduces communication by a factor of $H$. \citet{douillard2023diloco} introduced \emph{DiLoCo}, a low-communication variant of local SGD where each replica\footnote{A replica refers to a node, or pipeline-parallel set of nodes, containing a copy of model weights for DiLoCo training.} computes pseudo-gradients on its copy of the weights and data for $H$ steps, then the pseudo-gradients are averaged across the replicas in an all-reduce operation. \citet{douillard2025streaming} further optimized the algorithm to yield \emph{Streaming DiLoCo}, wherein subsets of the model parameters are synchronized in sequence, rather than all at once, and workers continue training during the parameter synchronization. By overlapping communication with compute, the system becomes compute-bound when $H$, the number of local steps between synchronizations, is sufficiently large. These advances progressively reduced the bandwidth required for distributed training. This research is extended in Decoupled DiLoCo \citep{douillard2026decoupled}, an asynchronous variant that tolerates partial hardware failures and heterogeneous chips within a single run.

\citet{erdil2024movement} theoretically model the effectiveness and limitations of large-scale model training subject to data movement bottlenecks, and \citet{erdilbesiroglu2024simulator} published an interactive simulator illustrating the model's prediction of training performance depending on the user's input. This work builds on theirs to simulate the case of multiple sites, and accommodates lower bandwidth and greater latency.

\citet{krys2025distributed} provided a taxonomy of governance risks from distributed and decentralized training, including proliferation of dangerous capabilities, decreased shutdownability of misaligned or misused models, and lack of oversight. However, they did not assess the feasibility of frontier-scale distributed training with small nodes, nor evaluate the effectiveness of countermeasures, which are analyzed below.

\citet{scher2025asi} proposed an international agreement that, if enacted, would restrict AI training in countries party to the agreement. The proposal would ban model pre-training over \flop{24} and fine-tuning over \flop{23}, with model training between \flop{22} and \flop{24} allowed but monitored. It would also require all clusters of chips with more computing power than the equivalent of 16 H100 GPUs to be registered and monitored. The proposal explicitly intends to make it difficult to perform distributed training across multiple sub-threshold sets of chips. However, recent advancements in distributed training algorithms enable exactly this type of violation: models larger than their banned threshold can be trained in a relatively short timeframe using nodes with processing rates of 16 H100-equivalents or less, as demonstrated herein. This paper proposes an amendment to \citet{scher2025asi}'s cluster registration requirements that makes governance evasion much more difficult and detectable.

\citet{sevillatroynikov2025power} assessed the feasibility of synchronizing frontier-scale datacenters to create even larger supercomputers, in order to enable larger training runs than those that can be done using only one datacenter. They concluded that this is technically feasible and would allow developers to bypass power grid limitations, illustrating an example of 23 large datacenters connected with 4,800 km of fiber optic cables and a total power draw of 10 GW. Finally, \citet{sevilla2025decentralized} conducted a literature review of distributed training techniques and collected a dataset of models trained using these techniques, then used the data to forecast future scaling of distributed training runs. This paper extends Sevilla and Troynikov's work to cases of smaller nodes, more replicas, more local steps, lower bandwidth, and greater latency.

\section{Methodology}
\label{sec:methodology}

\subsection{Threat Model}
\label{sec:threat-model}

Suppose a governing body regulates and requires reporting of model training above a compute threshold, and reporting of compute clusters above a performance threshold. Then a developer can avoid complying with the model development regulations by using clusters of hardware below the performance threshold and not reporting their model. The scenarios analyzed here involve a well-resourced AI developer attempting this strategy.

We investigate the feasibility of developing models similar to Llama 3.1-405B \citep{meta2024llama} (the largest published open-source model by training compute \citep{rahman2025models,epoch2026aimodels}, and above most regulatory compute thresholds) under very difficult conditions: consumer-grade internet with 100 Mbps bandwidth and 100 ms latency---thousands of times slower than inter-server connections in datacenters---and a limit of only 16 H100-equivalents of compute per node, as in \citet{scher2025asi}, the most restrictive compute governance proposal in the literature. Furthermore, we assume the governing body is actively searching for any signs of illicit distributed training, by monitoring internet traffic, power and water usage, satellite imagery, and thermal radiation, as well as conducting physical inspections of suspected undisclosed computing facilities. We assume the time available for training any model is limited to two years or less; our conclusions are robust to alternative assumptions; see Appendix~\ref{app:training-time} for justification and analysis.

Pretraining is the most communication-heavy and compute-intensive phase of model development; fine-tuning and RL are less difficult to adapt to a distributed hardware configuration, so we conservatively focus on pretraining because it is the most challenging phase for the evader \citep{krys2025distributed}. If the evader succeeds at pretraining a large model, it is relatively easy to fine-tune it to improve its reasoning, coding, and other capabilities.

\subsection{Hardware Selection}
\label{sec:hardware-selection}

The evader selects the most useful hardware that can be purchased in sufficient quantities and performs the specified training workload under the above constraints, in particular, the limit of 16 H100-equivalent compute per node. In this case, the optimal hardware is often the NVIDIA A100 80GB, which is inexpensive, exists in large quantities, and has a very large amount of accelerator memory relative to its computing performance. The NVIDIA GH200 is also an effective choice, with superior FLOPS/\$, support for FP8 numeric format, and up to 144 GB of high-bandwidth memory per card. Other hardware options, including Google and Huawei chips, are compared in Appendix~\ref{app:hardware}. Recent work \citep{ryabinin2023swarm,douillard2026decoupled} demonstrates that distributed training can mix hardware types within a training run without degrading ML performance, so in practice, evaders are not constrained to a single hardware type; however, in the scenarios modeled, heterogeneous hardware would not outperform uniform hardware.

\subsection{Efficiency Model}
\label{sec:efficiency-model}

The most important metrics calculated by the simulator for any training configuration are $\Clocal$, the local-equivalent compute, and $\Cquality$, the quality-adjusted compute:
\begin{equation}
  \Clocal = \Cthroughput \times \etalocal,
  \label{eq:clocal}
\end{equation}
where $\Cthroughput$ is the nominal compute throughput, the total number of operations done by the processors and applied to model training,\footnote{Compute throughput is calculated using the standard formula $C = 6ND$.} and $\etalocal$ is the distributed training inefficiency factor, a ratio between 0 and 1 that accounts for the relative inferiority of distributing hardware across multiple nodes and communicating over relatively slow internet connections, rather than aggregating the hardware in a single cluster where data can be transferred at hundreds or thousands of gigabits per second. Local-equivalent compute is calculated as nominal compute throughput times the distributed training inefficiency factor.

\begin{equation}
  \Cquality = \Clocal \times \chi.
  \label{eq:cquality}
\end{equation}
$\Cquality$ is quality-adjusted, local-equivalent FLOPs---that is, the amount of centralized, optimally allocated compute that would produce the same quality model. $\chi$ is a penalty applied to models that are over- or under-trained relative to Chinchilla optimality, explained below.

The inefficiency factor $\etalocal$ consists of three separate factors, or four if pipeline parallelism is enabled:
\begin{equation}
  \etalocal = \etaH \times \etacomp \times \etarep \times \etaact.
  \label{eq:eta}
\end{equation}

\textbf{Sync interval penalty} ($\etaH$) is the decrease in efficiency from having replicas do many inner steps before synchronizing with each other, resulting in replicas drifting apart and computing stale gradients. If $H=1$, then the distributed training operation is simply doing data-parallel training with no local SGD, so there is no penalty. For higher values of $H$, $\etaH$ decreases logarithmically, with scaling coefficients calibrated to experiments published by \citet{stich2019local}, \citet{douillard2023diloco}, and \citet{charles2025diloco}. This is a small factor in $\etalocal$, causing around 2--10\% loss in effective compute for the training runs modeled in this work. It is modeled as $\etaH = 1 - \alpha\log_{10}(H)$, with $\alpha$ decreasing with model size, per \citet{charles2025diloco}. For example, a model with 250B parameters has $\alpha \approx 0.05$; at $H=50$, $\etaH \approx 0.91$.

\textbf{Compression quality} ($\etacomp$) accounts for the loss in compute efficiency from compressing gradients to save on bandwidth. Compressing the gradients reduces the amount of communication required but makes the transmitted gradient slightly inaccurate. This can be mitigated by error feedback methods, such as error feedback accumulation, wherein the difference between the gradient calculated locally and the gradient sent with compression is stored and added onto the next update. The SparseLoCo algorithm, implemented in Covenant-72B, is a prominent example of this. Based on Covenant-72B's published results \citep{lidin2026covenant}, the simulator's default compression setting is $150\times$, with a corresponding $\etacomp = 0.99$.

\textbf{Replica divergence} ($\etarep$) is the loss in compute efficiency from averaging gradients collected from replicas trained on different subsets of data and exploring different areas of the loss landscape. In general, averaging gradients from many replicas is not optimal: for example, if two replicas descend down two parallel valleys in the loss landscape with a hill between them, the average of their individual gradients may place the models on the hill. Based on \citet{charles2025diloco}, this penalty scales inversely with model size: the more high-dimensional the loss surface, the more similar the gradient paths experienced by all replicas. This is the dominant factor in $\etalocal$ for most training runs modeled here, with $\etarep$ ranging between 0.15 and 0.90.

\subsection{Chinchilla Quality Adjustment}
\label{sec:chinchilla}

Distributed training can sometimes face other constraints not present in single-cluster training---most crucially, limited accelerator memory---and may therefore structure models suboptimally relative to what would be done locally. We define quality-adjusted compute $\Cquality$ as the local-equivalent compute times a Chinchilla-suboptimality penalty ($\chi$), accounting for the fact that a Chinchilla-optimal model would achieve equal quality with a smaller training compute budget \citep{hoffmann2022chinchilla,besiroglu2024chinchilla}.

Since single nodes have much less memory available than the hardware in total, the maximum model size that can fit into memory during training or inference is much smaller than if training were done in a single cluster. This causes distributed training operations to tend to over-train their models. Some amount of overtraining is desirable \citep{erdil2024inference}, but too much results in lower training compute efficiency.

\subsection{Pipeline-parallel DiLoCo}
\label{sec:pipeline-parallel}

The activation quality penalty ($\etaact$) is the loss in compute efficiency from compressing activations from the model's hidden layers when using pipeline parallelism to shard the model across nodes.

At large node counts, flat DiLoCo severely overtrains small models. Therefore, it becomes worthwhile to use pipeline parallelism across nodes, using groups of nodes to hold a larger model than can fit on any one of them by itself. This incurs an efficiency penalty because activations must be compressed and transmitted between nodes over the internet, but at very large compute scales, this produces a greater amount of Chinchilla-optimal-equivalent compute. To simulate the effects of activation compression and pipeline bubbles, the simulator introduces the factor $\etaact$ within $\etalocal$ when the hardware configuration scenario is set to use PP groups. Otherwise, this factor is not applied, or equivalently, $\etaact=1$ in the equation for $\etalocal$ above.

\section{Results}
\label{sec:results}

The simulator's predicted feasibility of training models at larger scales, based on the scaling behavior derived from the published ML literature, is shown in Table~\ref{tab:main-results}. The following compute thresholds could be exceeded by distributed training operations, even under the hardware restrictions proposed in \citet{scher2025asi}, within the allotted time limit.

\begin{strip}
\vspace{-0.10in}
\begin{center}
\captionof{table}{Minimum-cost distributed hardware configurations achieving the specified levels of training compute. Columns show the hardware configuration (from node types in Appendix~\ref{app:hardware}), model size, inner step count $H$, compute efficiency $\etalocal$, Chinchilla quality-adjustment factor $\chi$, overtraining ratio (OT) relative to Chinchilla-optimal, and upfront hardware cost. Costs are computed as (price per chip)$\times$(chips)$\times$(chip-to-server factor $1.64\times$)$\times$(server-to-cluster factor $1.23\times$), adapted from \citet{cottier2024costs}. Training modes include flat, hierarchical (Hier), and pipeline-parallel (PP) DiLoCo variants. Training compute of DeepSeek-V3 (\scinum{3.3}{24}) and Llama 3.1-405B (\scinum{3.8}{25}) reported by \citet{deepseek2024v3} and \citet{grattafiori2024llama}; GPT-4 (\scinum{2.1}{25}) and GPT-5 (\scinum{6.6}{25}) estimated by \citet{epoch2026aimodels}.}
\label{tab:main-results}
\scriptsize
\resizebox{\textwidth}{!}{%
\begin{tabular}{@{}llllllllllll@{}}
\toprule
Target & Node Config & Nodes & Mode & Model & $H$ & $\etalocal$ & $\Clocal$ & $\chi$ & $\Cquality$ & OT & Cost \\
\midrule
$10^{24}$ & \chip{16}{H100 FP8} & 2 & Flat & 91B & 18 & 0.7957 & \scinum{1.3}{24} & 0.9796 & \scinum{1.3}{24} & $1.3\times$ & \$1.6M \\
DeepSeek-V3 & \chip{16}{GH200 FP8} & 7 & Flat & 160B & 19 & 0.5973 & \scinum{3.4}{24} & 0.9635 & \scinum{3.3}{24} & $1.4\times$ & \$6.3M \\
$10^{25}$ & \chip{16}{GH200 FP8} & 34 & Flat & 160B & 19 & 0.3698 & \scinum{1.0}{25} & 0.6250 & \scinum{6.4}{24} & $7.0\times$ & \$30.7M \\
GPT-4 & \chip{16}{GH200 FP8} & 101 & Hier ($8\times 12$) & 160B & 19 & 0.2580 & \scinum{2.1}{25} & 0.3689 & \scinum{7.8}{24} & $21\times$ & \$91.3M \\
Llama 3.1-405B & \chip{50}{A100 80GB} & 625 & Hier ($10\times 62$) & 250B & 19 & 0.1524 & \scinum{3.8}{25} & 0.3214 & \scinum{1.2}{25} & $26\times$ & \$441.3M \\
GPT-5 & \chip{16}{H100 FP8} & 2,880 & PP ($2\times 1440$) & 180B & 3 & 0.4244 & \scinum{6.6}{25} & 0.2901 & \scinum{1.9}{25} & $31\times$ & \$2.32B \\
$10^{26}$ & \chip{16}{H100 FP8} & 4,706 & PP ($2\times 2353$) & 180B & 3 & 0.3934 & \scinum{1.0}{26} & 0.2123 & \scinum{2.1}{25} & $51\times$ & \$3.80B \\
\bottomrule
\end{tabular}%
}
\end{center}
\vspace{-0.10in}
\end{strip}

Based on the compute thresholds of \flop{24} in \citet{scher2025asi}, \flop{25} in the EU AI Act, and \flop{26} in California SB 53, these proposed or existing regulations could be evaded using \$1.6M, \$31M, and \$3.8B, respectively, of hardware arranged in clusters smaller than any proposed registration requirement. This poses a severe challenge for regulation and governance, which cannot be applied to models and hardware when the governing body cannot reliably know that they exist. This also makes it difficult to stop the proliferation of dangerous capabilities that emerge at this scale. 101 sub-monitoring nodes reach the local-equivalent compute of GPT-4, and 625 nodes reach the training compute of Llama 3.1-405B.

Above that level, hardware requirements increase sharply due to decreasing efficiency at high replica counts, though the increase is much more modest if larger nodes can be used.

Appendix~\ref{app:bandwidth} illustrates the cost to achieve \flop{25} of local-equivalent compute with different amounts of available network bandwidth. At 100 Mbps bandwidth, it costs about $3\times$ more to do training distributed relative to centralized, but it is not a substantial barrier to feasibility. Latency has no significant effect: transmission times at DiLoCo's sync volumes are greater than round-trip time by several orders of magnitude.

\section{Discussion}
\label{sec:discussion}

\subsection{Assumptions}
\label{sec:assumptions}

The simulator estimates evader capability using techniques demonstrated in published ML experiments, and extrapolates their performance to larger scales where the developer uses more parameters and compute. The extrapolations are typically pessimistic, assuming worse performance beyond the maximum scale where they have been tested. A full catalog of parameter values is provided in the accompanying documentation and the project repository. Actual distributed training performance may improve, especially with the application of new techniques. Distributed training is an active area of research, with development ongoing in major labs such as Google \citep{douillard2026decoupled}.

The simulator's most recent calibration test was Covenant-72B \citep{lidin2026covenant}, released after the simulator's initial development. Covenant confirmed compute-bound operation ($94.5\%$ time spent computing even without Streaming DiLoCo), and achieved $146\times$ compression with negligible quality loss---nearly an order of magnitude above the simulator's original $16\times$ default. We updated the default to $150\times$ to reflect this result. Covenant matched Llama-2-70B quality with substantially fewer training tokens, validating the simulator's predictions at the 72B scale and exceeding several of its conservative assumptions.

In the opposite direction, three of our modeling choices are generous to the evader. First, we assume the evader has access to enough high-quality pretraining data to train at \flop{26} and beyond without reusing data across multiple epochs. At the upper end of the configurations we model, this implies tens of trillions of unique tokens, a corpus larger than what the open-web frontier has so far been able to assemble. A data-constrained evader would need to repeat data, which incurs diminishing returns not modeled by the standard Chinchilla scaling law and would moderately reduce model quality at fixed compute. Second, the time-limit derivation assumes the evader is willing to train for years under treaty-restricted growth rates. The treaty's enforcement mechanisms---whistleblower programs, challenge inspections, financial intelligence---create cumulative detection probability that scales with operational duration, pushing evaders toward shorter runs at the cost of total compute. Even at a six-month duration rather than two years, however, the configurations analyzed exceed the relevant compute thresholds by wide margins, so this consideration moderately affects the quality of illicitly developed prohibited models, but not the overall feasibility of developing them. Third, we assume that all workers finish processing their batch and transmit their update at the same time. In reality, random hardware failures and network conditions cause some to report late or drop out, but these effects can be mitigated by techniques demonstrated in past work and do not affect the feasibility of distributed training under realistic network conditions and hardware reliability rates.

Evidence for each assumed value from published ML literature is provided in the accompanying simulator documentation (see Appendix~\ref{app:simulator-docs}). Discussion of uncertainties in scaling and how they affect the simulator's predictions is provided in Appendix~\ref{app:uncertainties}, and a review of techniques accounting for stragglers and unreliable workers is provided in Appendix~\ref{app:stragglers}.

\subsection{Evader Strategy}
\label{sec:evader-strategy}

When targeting high quantities of quality-adjusted compute $\Cquality$, the simulator penalizes configurations that train small models, because these are overtrained relative to Chinchilla-optimal, and a developer could theoretically create a model with the same loss using a smaller compute budget. However, in practice, developers prefer their models to be moderately overtrained. This is because the amount of inference compute used over the lifetime of any model is generally within one order of magnitude of the upfront training compute \citep{erdil2024inference}, and developers can spend some multiple of extra inference compute to get the same effect as a multiple more training compute \citep{villalobos2023trading}---if inference compute were much smaller than training compute, the developer would train a smaller model and scale up the compute used during each inference, achieving a cost saving while maintaining the same output quality. Therefore, to optimize the utility of the model over its deployed lifespan while minimizing compute spending, developers tend to deliberately over-train models. Therefore, the simulator's $\Cquality$ metric is excessively pessimistic for modest overtraining ratios.

In the current world, hardware and software are rapidly improving and AI investments are growing sharply. On the other hand, if AI development were tightly restricted as proposed by \citet{scher2025asi}, slower growth rates would mean existing hardware, software, and budgets remain at the frontier for longer, and developers could spend a much longer time training a state-of-the-art model. There are two other effects on training time from a highly restrictive governance scenario. First, ongoing enforcement and intelligence operations, as well as whistleblower incentives, are more effective against longer operations, so evaders would prefer to train faster in order to minimize the timeframe in which they can be caught. On the other hand, for any given amount of compute, a bigger operation that trains faster with more hardware and greater throughput will have more sites, visibility, and financial costs than another operation that reaches the same compute more slowly using less hardware. This effect encourages developers to train over longer periods. The overall effect is that if developers are trying to evade governance, some of them, including those training the most capable models, would likely attempt long-running training operations, up to the time limits identified in \citet{sevilla2022longest}.

\subsection{Countermeasures}
\label{sec:countermeasures}

Several countermeasures are potentially relevant and could help prevent developers from evading AI governance using distributed training. The most effective measures include chip tracking, whistleblowing programs, memory and compute limits on unregistered hardware, and traditional intelligence and law enforcement techniques. Others, such as capping bandwidth available from AI computing sites and monitoring internet traffic, are likely ineffective because they are too easily evaded by developers with sufficient determination and technical capabilities. A detailed assessment of the efficacy of various countermeasures is provided in Appendix~\ref{app:countermeasures}.

\subsection{The Simulator}
\label{sec:simulator}

We release the \href{https://intelligence.org/research/distributed-training-simulator}{simulator}, with open-source Python backend and an interactive web interface, designed to allow governance researchers to evaluate evasion scenarios under their own assumptions. Configurable parameters include hardware assumptions (GPU type, node count, local MFU), network conditions (bandwidth, latency), DiLoCo configuration (inner steps, compression ratio, and flat, hierarchical, or pipeline-parallel mode), training duration, and model size. The simulator distinguishes raw local-equivalent compute ($\Clocal$) from quality-adjusted compute ($\Cquality$), and supports optimistic, expected, and conservative scenarios for compression-quality factors so that users can examine the sensitivity of any particular conclusion to extrapolation uncertainty. The codebase, scaling-law calibration, literature review of past distributed training experiments, and relevant assumptions are documented in the \href{https://github.com/robirahman/miri-decentralized-training-report}{project repository}.

\section{Conclusion}
\label{sec:conclusion}

Distributed training can produce models above every existing or proposed compute threshold, and as capable as current frontier models, even when using hardware \emph{below} every proposed monitoring threshold. Configurations evaluated herein would violate \citet{scher2025asi} for under \$2M, the EU AI Act for \$31M, and SB 53 for \$3.8B worth of unmonitored hardware. This leaves an enforcement gap in current governance methods, if developers become determined to evade them. Currently, the increased costs are economically unpalatable, but if developers encounter regulatory bottlenecks in the future, they may be willing to pay the price. Given the pace of ongoing research, the enforcement gap identified here is actively growing, and methods for detecting it will become even more important.

Existing technical countermeasures are insufficient. However, a combination of compute- and memory-based chip tracking substantially raises evasion costs. Combining chip registries with whistleblower programs, challenge inspections, and conventional intelligence operations reestablishes monitorability of distributed training operations and allows AI governance to continue without loopholes.

\section*{Impact Statement}
\addcontentsline{toc}{section}{Impact Statement}
This paper presents work whose goal is to advance the field of technical governance of AI. The work identifies potential failures of governance and investigates solutions in order to mitigate both existential and prosaic risks.

\section*{LLM Usage Statement}
\addcontentsline{toc}{section}{LLM Usage Statement}
No AI-generated text is included in the final draft of this paper. LLMs were used for assistance in identifying relevant literature, fitting simulator parameters to published experiments, implementing the simulator scripts, creating the web simulator frontend, fact-checking claims made in this document, and converting it to \LaTeX.

\section*{Acknowledgements}
\addcontentsline{toc}{section}{Acknowledgements}
Thanks to Aaron Scher for extensive feedback, to Jaime Sevilla for past reviews of distributed training that inspired the simulator concept, and to Alex Beck for editing assistance.

\appendix

\section{Training Time Limit}
\label{app:training-time}

We assume that development for any model is subject to a practical time limit of 740 days, based on the theoretical model published by Epoch in \citet{sevilla2022longest} and using model inputs motivated by the compute governance proposals in \citet{scher2025asi}.

\citet{sevilla2022longest}'s model gives an upper bound for training time when accounting for hardware improvements, algorithmic and software improvements, and rising AI investments. The key insight is that, when hardware is improving over time, an overly long training run will be overtaken by a shorter training run that starts later but uses newer, faster hardware. Similarly, if funding is increasing over time, a long training run with a small budget will be overtaken by a shorter training run that starts later but has a larger budget. The formula for maximum training time is then:
\begin{equation}
  T = \frac{1}{g_H + g_S + g_I},
  \label{eq:training-time}
\end{equation}
where $g_H$ is the rate of improvement of hardware efficiency, $g_S$ is the rate of improvement of software, and $g_I$ is the growth rate of AI development spending.

In the current, unregulated environment, growth rates are $1.37\times$ per year for AI chip FLOPS/\$ \citep{epoch2026hardware}, $3\times$ per year in pretraining algorithmic efficiency \citep{ho2024algorithmic}, and $3.5\times$/year in frontier model training costs \citep{epoch2026trainingruns}, resulting in a training time bound of approximately 4.5 months.

Based on correspondence with the authors of \citet{scher2025asi}, we model the evader's time constraint based on growth rates of 6\% per year for hardware, 50\% per year for software, and 3\% per year for investment. Their governance proposal is intended to completely stop the advancement of AI algorithms and hardware, but in practice, a partial reduction is more realistic. This slower progress would allow time for AI alignment researchers to improve model safety before the development of transformative, superhuman artificial intelligence, but it also incentivizes evaders to spend more time doing illicit model development. These input values represent a reduction of over 75\% in the rate of hardware and software progress, and cut investment scaling all the way down to the background level of economic growth, yielding a time limit of 740 days or approximately two years.

With intermediate levels of regulations, the training time used by evaders would correspondingly fall in between these values. Using the slower growth rates and longer time limit is potentially generous to the evader, but even training times at the lower end of this range are long enough for the evader to successfully train models at near-frontier scale, as illustrated in the Results section above. Implications of long training runs on detectability of the evader by the governing body are addressed in the Discussion section.

\section{Hardware Configurations}
\label{app:hardware}

The simulator offers a variety of node configurations so that researchers can estimate the efficiency and cost of training models on different types of hardware. Presets include standard racks or pods produced by the hardware manufacturers, as well as nodes sized at or below the 16 H100-equivalent hardware reporting threshold proposed in \citet{scher2025asi}.

All performance figures are in PFLOPS and all memory figures are in GB.

\subsection{Standard Commercial Pods}
\label{app:commercial-pods}

\begin{table}[!ht]
\centering
\caption{NVIDIA preset pods.}
\label{tab:nvidia}
\resizebox{\columnwidth}{!}{%
\begin{tabular}{@{}lrrr@{}}
\toprule
Preset & Chips & Perf. (FP16) & Memory \\
\midrule
GH200 NVL32 & 32 & 31.68 & 4,608 \\
GB200 NVL72 & 72 & 162.0 & 13,824 \\
DGX H100 SuperPOD & 256 & 253.44 & 20,480 \\
DGX A100 SuperPOD & 1,120 & 349.44 & 89,600 \\
\bottomrule
\end{tabular}%
}
\end{table}

\begin{table}[!ht]
\centering
\caption{Huawei preset pod.}
\label{tab:huawei}
\resizebox{\columnwidth}{!}{%
\begin{tabular}{@{}lrrr@{}}
\toprule
Preset & Chips & Perf. (FP16) & Memory \\
\midrule
CloudMatrix 384 (910C) & 384 & 230.4 & 49,152 \\
\bottomrule
\end{tabular}%
}
\end{table}

\begin{table}[!ht]
\centering
\caption{Google preset pods.}
\label{tab:google}
\resizebox{\columnwidth}{!}{%
\begin{tabular}{@{}lrrr@{}}
\toprule
Preset & Chips & Perf. (BF16) & Memory \\
\midrule
TPU v4 pod & 4,096 & 1,126.4 & 131,072 \\
TPU v5e pod & 256 & 50.43 & 4,096 \\
TPU v5p pod & 8,960 & 4,112.64 & 851,200 \\
TPU v6e pod & 256 & 235.01 & 8,192 \\
\bottomrule
\end{tabular}%
}
\end{table}

\subsection{Nodes Under the 16 H100-Equivalent Threshold}
\label{app:subthreshold-nodes}

\begin{table}[!ht]
\centering
\caption{Sub-threshold node configurations, 16-bit arithmetic.}
\label{tab:subthreshold16}
\resizebox{\columnwidth}{!}{%
\begin{tabular}{@{}lrrrrr@{}}
\toprule
Hardware & Chips & Perf. (16-bit) & H100-eq. & Memory & Cost/chip (USD) \\
\midrule
\chip{50}{A100 80GB} & 50 & 15.600 & 15.76 & 4,000 & \$7,000 \\
\chip{49}{Ascend 910B} & 49 & 15.680 & 15.84 & 3,136 & \$16,000 \\
\chip{26}{Ascend 910C} & 26 & 15.600 & 15.76 & 3,328 & \$26,000 \\
\chip{57}{TPU v4} & 57 & 15.675 & 15.83 & 1,824 & \$12,000 \\
\chip{80}{TPU v5e} & 80 & 15.760 & 15.92 & 1,280 & \$6,000 \\
\chip{34}{TPU v5p} & 34 & 15.606 & 15.76 & 3,230 & \$20,000 \\
\chip{17}{TPU v6e} & 17 & 15.606 & 15.76 & 544 & \$25,000 \\
\bottomrule
\end{tabular}%
}
\end{table}

\begin{table}[!ht]
\centering
\caption{Sub-threshold node configurations with 8-bit FP support.}
\label{tab:subthreshold8}
\resizebox{\columnwidth}{!}{%
\begin{tabular}{@{}lrrrrr@{}}
\toprule
Hardware & Chips & Perf. (8-bit) & Perf. (16-bit) & Memory & Cost/chip (USD) \\
\midrule
\chip{16}{H100 SXM} & 16 & 31.68 & 15.84 & 1,280 & \$25,000 \\
\chip{16}{GH200} & 16 & 31.68 & 15.84 & 2,304 & \$28,000 \\
\chip{17}{TPU v6e FP8} & 17 & 31.21 & 15.61 & 544 & \$25,000 \\
\bottomrule
\end{tabular}%
}
\end{table}

\section{Bandwidth Sensitivity}
\label{app:bandwidth}

\begin{table*}[t]
\centering
\caption{Minimum-cost distributed hardware configurations that achieve \flop{25} of training compute with different amounts of available bandwidth. As of April 2026, Chinese average bandwidth is 207 Mbps down and 47 Mbps up; US average is 310 Mbps down and 57 Mbps up \citep{speedtest2026}.}
\label{tab:bandwidth}
\scriptsize
\resizebox{\textwidth}{!}{%
\begin{tabular}{@{}llllllllllll@{}}
\toprule
Bandwidth & Node Config & Nodes & Mode & Model & $H$ & $\etalocal$ & $\Clocal$ & $\chi$ & $\Cquality$ & OT & Cost \\
\midrule
10 Mbps & \chip{16}{GH200 FP8} & 3,082 & PP ($2\times 1541$) & 310B & 4 & 0.4234 & \scinum{1.0}{25} & 0.9465 & \scinum{9.5}{24} & $1.6\times$ & \$2.79B \\
30 Mbps & \chip{50}{A100 80GB} & 168 & Hier ($10\times 16$) & 250B & 61 & 0.1493 & \scinum{1.0}{25} & 0.6213 & \scinum{6.2}{24} & $7.0\times$ & \$118.6M \\
China avg & \chip{16}{GH200 FP8} & 38 & Flat & 160B & 25 & 0.3266 & \scinum{1.0}{25} & 0.5969 & \scinum{6.0}{24} & $7.8\times$ & \$34.3M \\
US avg & \chip{16}{GH200 FP8} & 34 & Flat & 160B & 20 & 0.3639 & \scinum{1.0}{25} & 0.6250 & \scinum{6.3}{24} & $7.0\times$ & \$30.7M \\
100 Mbps & \chip{16}{GH200 FP8} & 34 & Flat & 160B & 19 & 0.3698 & \scinum{1.0}{25} & 0.6250 & \scinum{6.4}{24} & $7.0\times$ & \$30.7M \\
300 Mbps & \chip{16}{H100 FP8} & 23 & Flat & 91B & 7 & 0.5391 & \scinum{1.0}{25} & 0.4507 & \scinum{4.5}{24} & $15\times$ & \$18.6M \\
1 Gbps & \chip{16}{H100 FP8} & 16 & Flat & 91B & 2 & 0.8160 & \scinum{1.1}{25} & 0.5370 & \scinum{5.7}{24} & $10\times$ & \$12.9M \\
\bottomrule
\end{tabular}%
}
\end{table*}

Table~\ref{tab:bandwidth} illustrates the cost to achieve \flop{25} of local-equivalent compute with different amounts of available network bandwidth. At 100 Mbps bandwidth, it costs about $3\times$ more to do training distributed relative to centralized, but distribution is not a substantial barrier to feasibility. Latency has no significant effect: transmission times at DiLoCo's sync volumes are greater than round-trip time by several orders of magnitude.

\section{Simulator and Modeling Documentation}
\label{app:simulator-docs}

A bibliography of past distributed training research, and comprehensive documentation of the simulator's mechanics, are available on GitHub.\footnote{Project repository: \url{https://github.com/robirahman/miri-decentralized-training-report}.} The bibliography includes published ML experiments whose data are used to estimate each parameter, and the documentation explains how parameters and outputs are estimated in the simulator backend.

\section{Principal Uncertainties}
\label{app:uncertainties}

Two structural uncertainties limit our ability to make precise quantitative claims at the upper end of the modeled range. The first is the empirical scaling of distributed training itself. The largest published DiLoCo-based run is Covenant-72B at 72.7B parameters and \scinum{4.8}{23}---roughly an order of magnitude fewer parameters, and two orders of magnitude less compute, than configurations we expect to be feasible in practice. The simulator's central efficiency factors---the sync-interval penalty coefficient, the replica divergence exponent, the per-boundary activation compression quality factor---are fit to experiments at $\leq 16$B parameters with $\leq 16$ replicas, and our conclusions extrapolate these by a factor of $10$--$300\times$ in model size and up to $250\times$ in replica count. PP-Group DiLoCo is particularly sensitive: at 8 pipeline stages, based on conservative values for loss degradation from activation compression, it compounds to $\etaact \approx 0.56$, reducing $\Cquality$ at 2,000 nodes by 49\% relative to the best-guess values estimated from existing experiments. The qualitative finding that distributed training crosses the compute thresholds in existing and proposed compute governance is robust, but there is substantial uncertainty in the precise values when the scale exceeds \flop{26}.

The second uncertainty relates to the parameters of the Chinchilla scaling law. Our quality-adjusted compute figures depend on the conversion from raw loss degradation to FLOP-equivalent penalty, which is governed by the scaling exponents published in \citet{hoffmann2022chinchilla}, fit to models with up to 280B parameters and \scinum{6}{23} of training compute. Crucially, this uncertainty is not directionally conservative: depending on the scaling behavior above this scale, distributed training could be less or more capable than our $\Cquality$ figures suggest and evasion may be less or more effective. Hardware costs, node counts, and qualitative feasibility assessments are unaffected because they depend on local-equivalent compute throughput rather than on the loss-to-FLOP conversion.

\section{Hardware Failures and Straggler Mitigation}
\label{app:stragglers}

The scenarios illustrated in this paper assume that all workers complete their batches at the same time. In reality, processors have some rate of random failure and not all traffic arrives simultaneously. If the entire distributed training network has to wait for every replica to calculate and send in its update, as in a naive single-program, multiple-data implementation, much time can be wasted by a few late or faulty workers. Several techniques exist to mitigate the effects of straggler nodes, such as aggregating and broadcasting the overall gradient update once some fraction, for example 90\%, of replicas have returned their pseudo-gradients. The late updates can be added to the next batch. SWARM demonstrates distributed training with poorly connected, unreliable devices, adaptively rebalancing groups to bypass slowdowns from lost nodes \citep{ryabinin2023swarm}. Decoupled DiLoCo runs learners independently and aggregates once a quorum has been reached, weighting the individual updates based on the number of tokens processed and number of steps completed by each worker \citep{douillard2026decoupled}.

Ultimately, stragglers do not affect the conclusions presented below at realistic rates of hardware failures \citep{grattafiori2024llama} and under typical network conditions. For example, Meta reported 419 hardware failures while training Llama 3.1-405B on 16,384 H100 GPUs over a period of 54 days, which is one hardware failure per 50,000 GPU-hours \citep{grattafiori2024llama}. Different assumptions such as unreliable hardware or networking, and mitigations such as aggregating upon reaching a threshold of workers, can be explored using the simulator.

\section{Countermeasure Analysis}
\label{app:countermeasures}

This section explores potential countermeasures to prevent governance evasion through distributed training, starting with the least viable options and moving on to the most effective measures.

\begin{table*}[t]
\centering
\caption{Countermeasures and their assessed effectiveness if implemented, along with qualitative societal burden on parties not conducting distributed training.}
\label{tab:countermeasures}
\begin{tabular}{@{}lllll@{}}
\toprule
Countermeasure & vs. Non-State & vs. State & Burden & Recommended? \\
\midrule
Bandwidth caps & High & Low & Extreme & No \\
ISP traffic monitoring & Very low & Very low & Medium & No \\
Chip registry with location verification & High & Medium & Low & Yes \\
Whistleblower programs & High & Low & Very low & Yes \\
Memory threshold (1,280 GB) & Medium & Medium & Low & Yes \\
Challenge inspections & Medium & Medium & Low & Yes \\
\bottomrule
\end{tabular}
\end{table*}

\subsection{Bandwidth Caps}
\label{app:bandwidth-caps}

As illustrated by Table~\ref{tab:bandwidth}, reducing the evader's bandwidth would substantially hinder the training throughput and increase the cost of evasion, but it is unfortunately infeasible to prevent evaders from getting fast internet connections. When each node contains hundreds of thousands of dollars worth of hardware, the cost of above-average internet is negligible by comparison, and there are too many broadband customers to audit them all. Nodes could be interspersed with residential facilities, and restricting consumer bandwidth to far below current levels would be an extreme, politically unpalatable measure. DiLoCo remains workable even at lower throughput levels, using greater pseudo-gradient compression and inner step counts.

\subsection{Internet Traffic Monitoring}
\label{app:traffic-monitoring}

Today, the vast majority of internet traffic is encrypted, so any efforts at detecting distributed training by monitoring internet traffic would have to rely on analysis of the volume and routing, not on the content. DiLoCo-based training achieves $>10^{25}$ FLOP with $\leq 16$ H100-eq per node and less than 40 Mbps of traffic. This is substantially less than the average US household internet speed, 310 Mbps down and 57 Mbps up \citep{speedtest2026}. If the sites have flexibility to go up to the average connection speed, the traffic can be disguised into almost any irregular pattern. Traffic shaping methods include jitter, variable-rate streaming, and intermittent pauses. All of these can be done while keeping the training hardware compute-bound and therefore having no detrimental effects on performance. And by routing through VPNs and parameter servers, they can increase the difficulty of tracing individual nodes to the rest of the training network. More research is needed to determine if any training detection methods are feasible, but the requirements of this problem appear to favor evaders.

\subsection{Chip Tracking}
\label{app:chip-tracking}

A chip registry with mandatory reporting of possession, transfer, and decommissioning was proposed by \citet{fist2023smuggling} in the context of export controls and echoed in \citet{scher2025asi} in the context of international cooperation on AI regulation. This would be a potent countermeasure against illicit AI development in general, but is especially effective against distributed training. Since a distributed configuration is less efficient than a centralized cluster, it requires more total hardware, so a registry that tracks a large fraction of existing compute is hard to evade in this way. \citet{baker2025verification} extend this framing, combining compute accounting (registry-based ownership tracking) with hardware-enabled mechanisms (HEMs).

Two HEM proposals are particularly relevant. \emph{Delay-based location verification}, proposed by \citet{brass2024location}, uses cryptographic challenge-response protocols between chips and landmark servers to bound each chip's physical location via speed-of-light delay. If deployed at manufacture on a sufficiently high percentage of AI chips, location verification would preclude distributed training. If the chips reported their locations, the governing body would be able to track down the nodes, and if a large quantity of chips went dark to avoid reporting, this would alert the authorities to an evasion attempt. An evader could attempt to use unregistered or untracked chips, but this would require large-scale smuggling or fabrication. \emph{Offline licensing} \citep{kulp2024hardware} requires chips to hold time-limited licenses, periodically renewed with the governing authority, to operate. The companion \emph{Fixed set} mechanism restricts high-bandwidth communication to pre-authorized pods of chips, directly blocking the inter-node communication DiLoCo depends on. However, these mechanisms do not exist on current hardware and cannot be retrofitted without certain preexisting firmware, so they cannot catch evaders using many types of already-deployed chips, a gap that persists for the useful lifetime of the current installed base. The effectiveness of a future registry is also diminished by developments that allow distributed training networks to effectively utilize heterogeneous hardware \citep{ryabinin2023swarm,douillard2026decoupled}, which would allow evaders to illicitly supplement new, tracked hardware with older, unregistered hardware to augment the throughput of a training run, pushing it above a compute threshold without the regulator's awareness.

\subsection{Whistleblowing Programs}
\label{app:whistleblowing}

Whistleblower programs are a simple, low-burden countermeasure that is well-suited to the distributed training threat model, because distributed evasion is more personnel-intensive than centralized training. An evader operating hundreds or thousands of nodes to train a frontier-scale model needs a much larger team to install and operate the network than to build one cluster in a datacenter. Procurement, installation, and ongoing maintenance across hundreds of locations requires a substantial workforce, and every person in that workforce is a potential reporter. Unlike centralized training, which can in principle be conducted by a small team inside a single secure facility, a distributed operation has an attack surface proportional to its node count.

\citet{baker2025verification} identify whistleblower programs as one of the six independent verification layers needed for robust AI agreement enforcement, and \citet{scher2025asi} explicitly include whistleblowing in their enforcement regime alongside supply-chain tracking and challenge inspections. The design template is well-established: the SEC Whistleblower Program, created under the Dodd-Frank Act, offers reporters 10--30\% of any resulting penalty along with anti-retaliation protections, and has recovered billions of dollars since its inception. The bipartisan Stop Stealing Our Chips Act (SSOCA), introduced in both chambers of Congress in 2025 and passed by the Senate as of 2026 \citep{rocco2026ssoca}, would apply this model directly to AI chip export violations at the Bureau of Industry and Security, with financial incentives and anonymous reporting portals \citep{grunewald2025whistleblower}. A parallel program covering unauthorized training runs, not only chip diversion, would extend this same mechanism to close the governance loophole modeled in this paper.

Whistleblower programs have two properties that make them particularly valuable in combination with the other countermeasures discussed above. First, they require no new hardware or chip design changes and therefore work immediately against the current installed base, unlike location verification and offline licensing. Second, their effectiveness scales with the evader's operational complexity, so memory caps and chip registries that force evaders to operate more nodes also amplify whistleblowing's reach. The main limitation is jurisdictional: programs administered by a single state cannot easily reach insiders at state-actor operations abroad, which is why whistleblowing must be combined with the other layers rather than relied on in isolation.

\subsection{Memory Limits}
\label{app:memory-limits}

Every cluster-reporting threshold proposed or enacted to date defines cluster size in compute terms alone. Executive Order 14110 and its implementing BIS proposed rule (2023--2024) required reporting of computing clusters with aggregate performance exceeding $10^{20}$ OP/s connected at over 300 Gbit/s \citep{whitehouse2023eo14110}. \citet{scher2025asi} would require registration of any Covered Chip Cluster (CCC) with aggregate effective computing capacity above 16 H100-equivalents (15,840 TFLOP/s), a threshold three orders of magnitude lower than EO 14110 but likewise defined solely by compute throughput.

This shared framing creates an exploitable asymmetry: chips with large accelerator memory relative to their compute, such as the A100 80GB, can hold substantially larger models per node while remaining below any compute-only threshold. A sub-threshold node of 50 A100 80GBs, for example, contains 4,000 GB of HBM---more than three times the memory of the reference 16-H100 node, while having a similar processing rate. As demonstrated in Table~\ref{tab:main-results}, such nodes are favorable choices for evaders, precisely because a high memory-to-compute ratio minimizes the communication and overtraining penalties that dominate distributed training efficiency. We therefore recommend that \citet{scher2025asi} revise their CCC definition to incorporate an accelerator memory threshold of 1,280 GB HBM, matching the memory of 16 H100s. Clusters should be covered if they exceed the capacity of 16 H100s in either memory or computational throughput.

This amendment forces the evader into one of two inefficient regimes. They can either shrink the model size to fit within the memory of each node---which, if trained to very large compute budgets, results in severe overtraining and a significant quality degradation ($\chi < 1$) relative to a Chinchilla-optimal model---or they can shard models across nodes using pipeline parallelism, which triggers activation-compression inefficiencies ($\etaact < 1$) and substantially increases the number of nodes required. Under the configurations in Table~\ref{tab:main-results}, imposing the memory limit reduces the maximum unregistered per-node model from 240B to roughly 91B parameters, forces the evader to operate approximately five times as many nodes for equivalent model quality, and raises the cost of illicitly developing a \flop{25} model by roughly 50\%.

The amendment imposes minimal collateral burden. Standard research workstations (4--8 GPUs), DGX-class servers (640 GB), cloud inference instances, and rendering farms using GDDR rather than HBM memory all fall well below the threshold. Only clusters equipped with 4,000 GB of HBM or more---usually 17 or more specialized AI chips---require registration, which are precisely the configurations that are of concern for AI governance. Notably, the memory threshold does not make evasion impossible; it makes evasion \emph{expensive and operationally complex}, which amplifies the effectiveness of the other countermeasures discussed above: a larger node count means more procurement, more personnel, more facilities, and therefore more surface for detection by chip tracking, whistleblower reports, and conventional intelligence operations.

\subsection{Conventional Intelligence Operations}
\label{app:conventional-intelligence}

The countermeasures discussed above---chip tracking, whistleblowing, and memory-threshold registration---are technical and regulatory instruments specific to compute governance. They are complemented by the same conventional intelligence tools that states routinely apply to other forms of illicit economic activity, and which \citet{baker2025verification} include as their sixth verification layer alongside the more AI-specific mechanisms.

Financial and procurement intelligence are very well-suited to uncovering distributed training. The configurations in Table~\ref{tab:main-results} require millions to billions of dollars of AI hardware plus corresponding expenditure on facilities and personnel---roughly \$441M to illicitly develop a model at the scale of Llama 3.1-405B under \citet{scher2025asi}'s constraints, and several billion at the scale of \flop{26}. Procurement of this magnitude is easily detectable when done by private entities in market economies. Just as financial structuring, breaking transactions into sub-threshold amounts to evade Bank Secrecy Act reporting, is detectable by anti-money-laundering enforcement, the governing body could audit organizations suspected of making undisclosed compute-related purchases.

Challenge inspections are on-site inspections conducted on short notice, and can be performed at facilities with known or suspected relevance to international nonproliferation treaties. They were conducted bilaterally 18 times per year by the United States and Russia from 2011 to 2022 on each other's nuclear facilities, establishing a clear precedent for their usage in international AI governance \citep{wasil2024verification}. They are named by \citet{scher2025asi} as a component of governance verification and are particularly effective against distributed training because discovery and physical inspection of even a single node can yield evidence that exposes the broader operation. However, asynchronous training architectures such as \citet{douillard2026decoupled} tolerate the loss of individual nodes or groups without interrupting the run, so inspections should be coordinated across the suspected network rather than executed piecemeal. Human intelligence and national technical means round out the stack. None of these tools individually defeat a determined state-scale evader, but combined with the preceding methods, they restore the ability to monitor and reveal covert adversaries.

\bibliographystyle{icml2026}
\bibliography{references}

\end{document}